\documentclass[aps,showpacs,nofootinbib,prc,twocolumn]{revtex4-1}
\usepackage{amsmath}
\usepackage{amssymb}
\usepackage{amsfonts}
\usepackage{graphicx}
\usepackage{bbold}
\usepackage{bm}
\usepackage{bm}
\usepackage{cancel}
\usepackage{times,float}
\usepackage{graphicx}
\usepackage[usenames,dvipsnames,svgnames]{xcolor}
\usepackage{slashed}
\usepackage{hyperref}
\hypersetup{colorlinks=true, linkcolor=Black, citecolor=Black,urlcolor=Black}
\usepackage{multirow}
\usepackage{ulem}

\newcommand{\NREC}{N_{\Lambda \,\mbox{\tiny{REC}}}}
\newcommand{\NQGP}{N_{\Lambda \,\mbox{\tiny{QGP}}}}
\newcommand{\NbarREC}{N_{\overline{\Lambda} \,\mbox{\tiny{REC}}}}
\newcommand{\NbarQGP}{N_{\overline{\Lambda} \,\mbox{\tiny{QGP}}}}

\begin{document}

\title{$\Lambda$ and $\bar{\Lambda}$ global polarization from the core-corona model.}

\author{Alejandro Ayala$^{1,2,3}$}
\author{Isabel Dom\'inguez$^4$}
\author{Ivonne Maldonado$^5$}
\author{Maria Elena Tejeda-Yeomans$^{6,7}$}
  \address{
	$^1$Instituto de Ciencias
	Nucleares, Universidad Nacional Aut\'onoma de M\'exico, Apartado
	Postal 70-543, CdMx 04510,
	Mexico.\\
	$^2$Centre for Theoretical and Mathematical Physics, and Department of Physics,
	University of Cape Town, Rondebosch 7700, South Africa.\\
	$^3$Departamento de F\'\i sica, Universidade Federal de Santa Maria, Santa Maria, RS 97105-900, Brazil.\\
	$^4$Facultad de Ciencias F\'isico-Matem\'aticas, Universidad Aut\'onoma de Sinaloa,
	Avenida de las Am\'ericas y Boulevard Universitarios, Ciudad Universitaria,
	C.P. 80000, Culiac\'an, Sinaloa, Mexico.\\
	$^5$Joint Institute for Nuclear Research, Dubna, 141980 Russia.\\
	$^6$Facultad de Ciencias - CUICBAS, Universidad de Colima, Bernal D\'iaz del Castillo No. 340, Col. Villas San Sebasti\'an, 28045 Colima, Mexico.\\
	$^7$Perimeter Institute for Theoretical Physics, 31 Caroline Street North Waterloo, Ontario N2L 2Y5, Canada.
}


\begin{abstract}
We report on work aimed to describe the $\Lambda$ and $\bar{\Lambda}$ global polarizations in a heavy-ion collision environment using the core-corona model, where the source of these hyperons is a high-density core and a less dense corona. We show that the overall properties of the polarization excitation functions can be linked to the relative abundance of $\Lambda$s and $\bar{\Lambda}$s coming from the core versus those coming from the corona. Both global polarizations
peak at collision energies $\sqrt{s_{NN}} \lesssim 10$ GeV. The exact positions and heights of these peaks depend not only on a reversal of relative abundances with collision energy, but also on the centrality class, both related to the QGP volume and lifetime. 
\end{abstract}


\maketitle

\section{Introduction}
Hyperon polarization is a key observable in high energy physics which allows to monitor the properties of spin in a reaction. $\Lambda$ global polarization has become an important observable in the study of the hot and dense matter created in heavy-ion collisions, crucial to determine some of the fundamental properties of the Quark-Gluon Plasma (QGP), such as vorticity, viscosity and flow. It is also one of the observables that can provide a guide to study criticality in the  phase diagram of strongly interacting matter~\cite{Becattini_rev,Becattini_2022,Alzhrani_Ryu_Shen_2022,Singh_Alam_2021,Wu_Yi_Qin_Pu_2022}. 

Recent  measurements of this global polarization in non-central collisions has been reported by the STAR Collaboration~\cite{STAR:2017ckg}. The measurements are consistent with the emergence of an overall large angular momentum $\sim 10^5 \hbar$ in the hot and dense matter produced in the reaction which is responsible for shear forces that in turn produce vorticity. When this vorticity couples to the spins of the QGP constituents, the latter align with the overall angular momentum. Measurements of $\Lambda$ polarization thus provide a quantitative insight into the produced vorticity of the QGP. The results also show that $\Lambda$ and $\bar{\Lambda}$ global polarization increases, as the energy of the collision decreases, and that this effect is larger for $\bar{\Lambda}$ than for $\Lambda$.

The heavy-ion community is still carrying out systematic simulations, building models and putting forward predictions to understand this differentiated behaviour at low energies. For instance, a recent work using UrQMD~\cite{Deng:2020ygd} shows an energy dependence of the kinematic and thermal vorticities, with a maximum at low energies ($2.5 - 4.0$ GeV). This energy domain is covered by several current experiments such as HADES ($2 - 2.4$ GeV), STAR-FXT ($3 - 7.7$ GeV) and STAR-BES II ($7.7 - 19$ GeV), and will also be explored in the future by the MPD at NICA ($4 - 11$ GeV)~\cite{MPD:2022qhn,Meehan:2017cum, HADES:2009aat,Yang:2019bjr}. Understanding the properties of these polarizations will help to better characterize the properties of matter produced in this energy domain. Different approaches to the study of hyperon global polarization have been put forward~\cite{Ivanov:2020udj,Karpenko:2016jyx,Sun:2017xhx,Deng:2021miw,Ivanov:2022ble,Xie:2019wxz,Csernai:2018yok}, mainly focusing on the behavior of $\Lambda$ polarization at collision energies $\gtrsim 7$ GeV, where the difference with the $\bar{\Lambda}$ polarization is not significant. In this work we report on an approach we developed using a core-corona model, namely, a two component model for global $\Lambda$ and $\bar{\Lambda}$ polarization~\cite{Ayala:2020soy,Ayala:2021xrn}. The approach provides a differentiated behavior of these polarizations at low collision energies for non-central collisions, due to the interplay between the relative abundances of the hyperons coming from the core and the corona.

\section{Core-Corona Model}

\begin{figure}
\centering \includegraphics[scale=0.3]{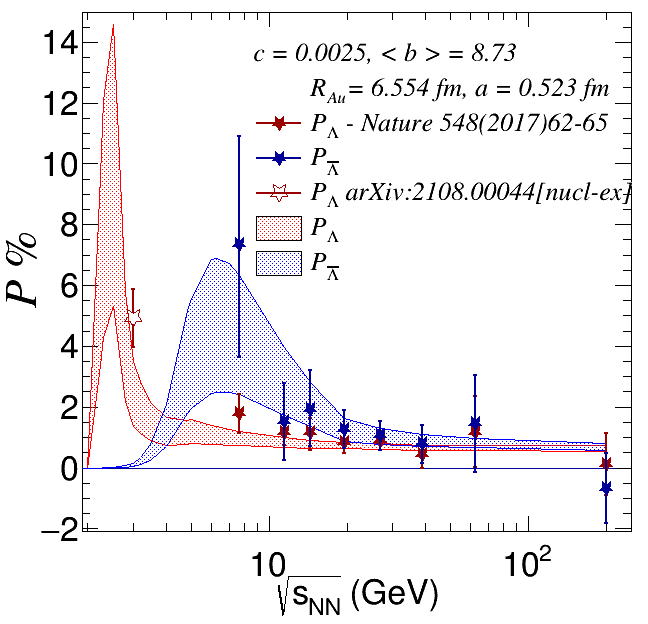}
 \caption{Polarization as a function of the collision energy for a centrality range 20\% - 50\%, compared with STAR BES data. Shaded areas correspond to the region delimited by the fits to the QGP vo\-lume and lifetime~\cite{Ayala:2021xrn}.}
 \label{Fig:original}
\end{figure}

The model assumes that in non-central collisions, $\Lambda$ and $\bar{\Lambda}$ hyperons can be produced from different density zones within the interaction region: core or corona, where their average global polarization can be obtained from 
\begin{eqnarray}
\mathcal{P}^\Lambda=\frac{z\frac{
N_{\Lambda\ {\mbox{\tiny{QGP}}}} }{N_{\Lambda\ {\mbox{\tiny{REC}}}}}}{ \left( 1 + \frac{N_{\Lambda\ {\mbox{\tiny{QGP}}}}}{N_{\Lambda\ {\mbox{\tiny{REC}}}}}\right)},\,\,\,
\mathcal{P}^{\overline{\Lambda}}=\frac{\bar{z}\left(\frac{w'}{w}\right)
\frac{N_{\Lambda\ {\mbox{\tiny{QGP}}}} }{N_{\Lambda\ {\mbox{\tiny{REC}}}}}}{ \left( 1 + 
\left(\frac{w'}{w}\right)
\frac{N_{\Lambda\ {\mbox{\tiny{QGP}}}}}{N_{\Lambda\ {\mbox{\tiny{REC}}}}}\right)}.
\label{eq1-mod}
\end{eqnarray}
\noindent as functions of the number of $\Lambda$s produced in the core $\NQGP$ and in the corona $\NREC$, the ratios of the number of $\bar{\Lambda}$s to $\Lambda$s, $w'=\NbarQGP/\NQGP$ and $w=\NbarREC/\NREC$ in the core and the corona, and the intrinsic polarizations $z$ and $\bar{z}$ for $\Lambda$ and $\bar{\Lambda}$ respectively. As can be seen in Fig. \ref{Fig:original}, the results obtained with this model are in agreement with data for a wide range of collision energies, and also predict that $\Lambda$ and $\bar{\Lambda}$ polarizations peak at different collision energies $\sqrt{s_{NN}} \lesssim 10$ GeV~\cite{Ayala:2021xrn}. Equations~(\ref{eq1-mod}) assume that the global polarizations have contributions from $\Lambda$s and $\bar{\Lambda}$s created through different mechanisms in the core (QGP) and in the corona (REC): coalescence type processes in the former, recombination of a di-(anti)quark with a (anti)quark, in the latter. In the corona, $\Lambda$ and $\bar{\Lambda}$ production happens by means of N + N reactions.

 In this environment, the ratio $w=\NbarREC/\NREC$ can be obtained from a fit to p + p data, which provides a good description above threshold production, $\sqrt{s} > 4.1$ GeV, through p + p $\to$ p + p + $\Lambda + \bar{\Lambda}$ reactions. In the core, QGP processes make it equally as easy to produce $\Lambda$ and $\bar{\Lambda}$, given that in this region quarks and anti-quarks are freely available such that $(\bar{u} , \bar{d}, \bar{s})$ can find each other as easy as $(u, d, s)$. At HADES, NICA and RHIC energies, the chemical potential and temperature of the system created is crucial to determine the properties of the core. 

Having this in mind, we calculate the ratio $w'=\NbarQGP/\NQGP$ along the freeze-out trajectory, using
equilibrium distributions which are parameterized in terms of the chemical potential, the temperature and the collision energy~\cite{Randrup:2006nr}. The $\Lambda$ and $\bar{\Lambda}$ intrinsic polarizations, $z$ and $\bar{z}$ respectively, quantify the relative number of hyperons with spin aligned in opposite directions with respect to the total number of hyperons ($z N_{\Lambda} = N^{\uparrow}_{\Lambda} - N^{\downarrow}_{\Lambda}, \bar{z} N_{\bar\Lambda} = N^{\uparrow}_{\bar\Lambda} - N^{\downarrow}_{\bar\Lambda}$) both in the core and in the corona. Since reactions in cold nuclear matter are less efficient to couple spin with angular momentum, we focus on the intrinsic polarizations for both $\Lambda$ and $\bar{\Lambda}$ in the core. $z$ and $\bar{z}$ in the core are computed from a field theoretical model for the alignment between quark spin and vorticity~\cite{Ayala:2020ndx,Ayala:2019iin,Ayala:2020soy}. 

Finally, we estimate the number of $\Lambda$s produced in each region ($\NQGP, \NREC$) introducing a critical density of participants, above (below) which, the QGP is (is not) formed. The number of $\Lambda$s from the core is proportional to the number of participant nucleons in the collision above this critical value. The density of participants is given in terms of the thickness functions of the colliding system and the collision energy-dependent N + N cross-section. The thickness function, in turn, comes from the nuclear density Woods-Saxon profile. 

Since the ratio of the number of $\bar{\Lambda}$s to $\Lambda$s coming from the corona is less than one ($w=\NbarREC/\NREC < 1$), it amplifies the $\bar{\Lambda}$ global polarization, making it larger than the $\Lambda$ global polarization, in spite of the intrinsic $\Lambda$ polarization being larger than the intrinsic polarization of $\bar{\Lambda}$ ($z > \bar{z}$). This happens for collisions with intermediate to large impact parameters, which correspond to the kind of collisions that favor the development of a larger thermal vorticity.

\section{Core-corona vs. recent data}

\begin{figure}
 \centering \includegraphics[scale=0.3]{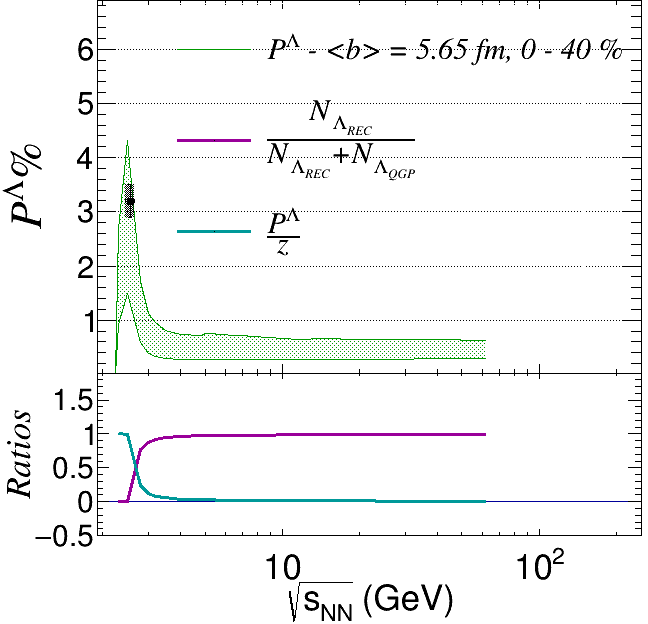}
\centering \includegraphics[scale=0.3]{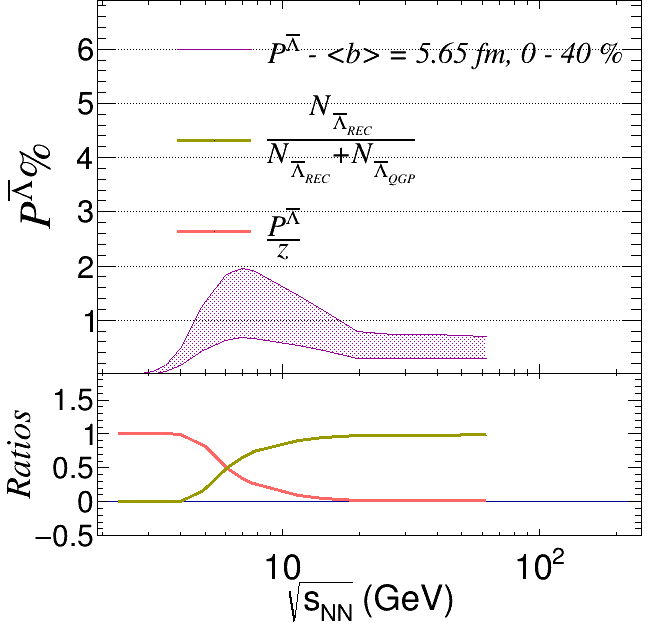}
 \caption{$\Lambda$ (upper panel) and $\bar{\Lambda}$ (lower panel) global polarization as a function of centrality calculated for Ag+Ag collisions. The pre\-li\-minary data point in the upper panel is from the HADES Collaboration~\cite{Kornas:2022cbl} measured for Ag+Ag collisions at $\sqrt{s_{NN}}=2.55$ GeV. As in the case for Au+Au, the $\bar{\Lambda}$ polarization~\cite{Ayala:2020soy,Ayala:2021xrn} increases faster than the $\Lambda$ polarization and both peak at a similar collision energy.}
 \label{Fig:AgAg2}
\end{figure}


The core-corona model that we have developed, describes well the behaviour of the hyperon average global polarization as a function of the collision energy for non-central collisions Au+Au, as reported in Refs.~\cite{Ayala:2020soy,Ayala:2021xrn}. However for the description of smaller systems, the model shows limitations. A comparison with preliminary results for hyperon global polarization in Ag+Ag collisions at $\sqrt{s_{NN}} = 2.55$ GeV, in the $10-40$\% centrality class~\cite{Kornas:2022cbl}, shows that the model predicts zero polarization. This feature is due to the lack of $\Lambda$s produced in the core, given that the critical density, $n_c$, required for QGP formation used in our calculation, is not achieved for collisions in this centrality class. To remedy this situation, we can account for the $\Lambda$s produced in the extended centrality range $0-40$\%. The behaviour of the polarization as a function of energy is then similar to the Au+Au case, as is shown in the upper panel of Fig.~\ref{Fig:AgAg2}. 

We observe that when changing the centrality interval, to include more central collisions, the model produces more $\Lambda$s in the core. This in turn modifies the magnitude of the average global polarization which also increases with the number of $\Lambda$s in the core. The lower panel in Fig.~\ref{Fig:AgAg2} shows the average global polarization for $\bar{\Lambda}$ as a function of the collision energy for Ag+Ag collisions. We observe that as the energy of the collision decreases, the polarization starts increasing earlier, peaking at a larger value than the corresponding case shown in the upper panel in Fig.~\ref{Fig:AgAg2}. The $\Lambda$ global polarization, as a function of impact parameter, shows an increasing behaviour up to $b \simeq 6$ fm, where $\NREC \sim \NQGP$, to then rapidly decrease to zero for $b \simeq 7.5$ fm. It turns out that this impact parameter corresponds to a situation in which the collision barely achieves the critical density $n_c$ to produce the QGP. 

\begin{figure}
 \centering \includegraphics[scale=0.3]{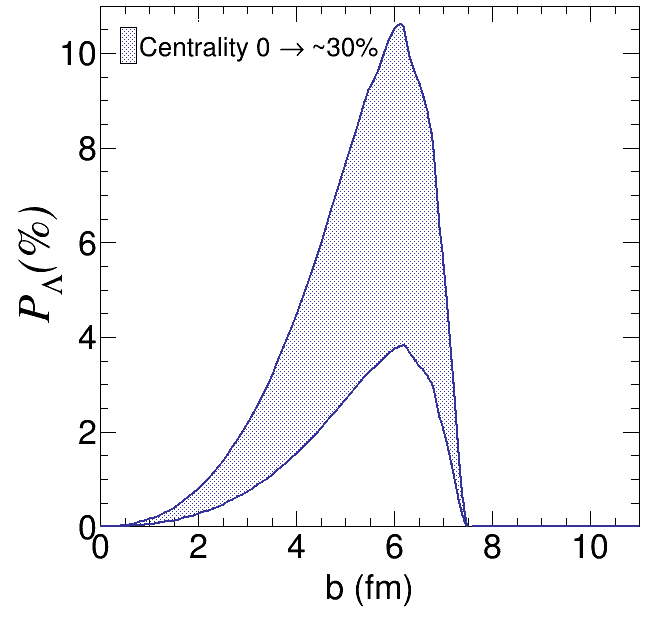}
\centering \includegraphics[scale=0.3]{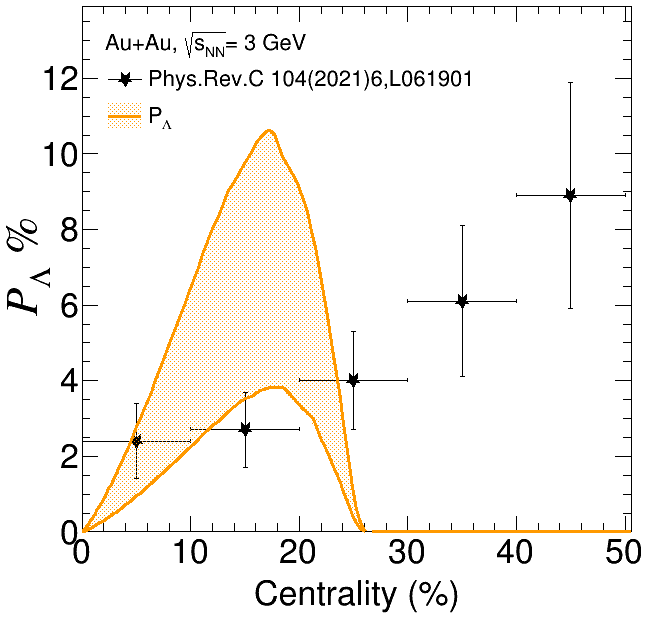}
 \caption{$\Lambda$ global polarization as a function of impact parameter (upper panel) and centrality (lower panel) calculated for Au+Au co\-llisions at $\sqrt{s_{NN}} = 3$ GeV. The lower panel shows data from the STAR Collaboration~\cite{STAR:2021beb}.}
 \label{Fig:polcent}
\end{figure}

The upper panel of Fig.~\ref{Fig:polcent} shows the $\Lambda$ global polarization as a function of impact parameter for Au+Au collisions at $\sqrt{s_{NN}}= 3$ GeV that can be compared with the most recent results on Au+Au collisions at $\sqrt{s_{NN}}= 3$ GeV from STAR~\cite{STAR:2021beb}. This is shown in the lower panel of Fig.~\ref{Fig:polcent}.

We observe a good agreement between model and data, for the two most central bins. The model shows a maximum for an impact parameter $b \simeq 6$ fm close to the value at which the critical density to produce the QGP is barely achieved, to then become zero. Notice however the increasing trend of data with centrality. We are aware that in the model this effect comes, in part, from the assumed sharp boundary of a critical density of participants $n_c$, between the core and corona regions. In our calculation, this was implemented by means of a step function, $\theta(n_p - n_c)$, in the density of participants $n_p$~\cite{Ayala:2020soy,Ayala:2021xrn}. A possible venue for improvement is to soften this criterion by using instead a smooth function,
\begin{equation}
f(n_p - n_c) = \frac{1}{1+2 e^{-2k(n_p - n_c)}},
    \label{eq:notheta}
\end{equation}
 with $k$ such that $f(n_p -n_c) \rightarrow \theta(n_p - n_c)$ as $k \rightarrow \infty$.  

\begin{figure}
 \centering \includegraphics[scale=0.3]{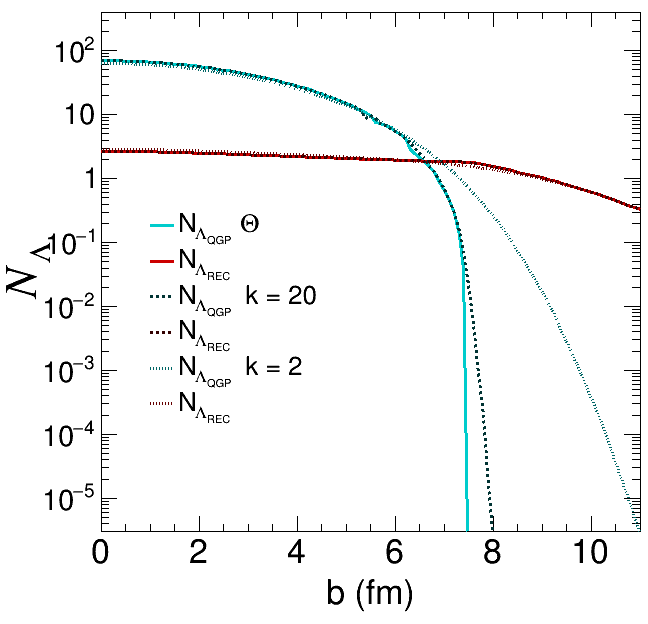}
 \caption{ Number of $\Lambda$s calculated with a step function (continuous lines), compared with the ones calculated with Eq.~(\ref{eq:notheta}) for two di\-ffe\-rent parameters (dashed lines). We observe that for $k=2$ it is possible to obtain $\Lambda$s from the core for larger impact parameter va\-lues.}
 \label{Fig:Nqgptheta}
\end{figure}


\begin{figure}
\centering \includegraphics[scale=0.3]{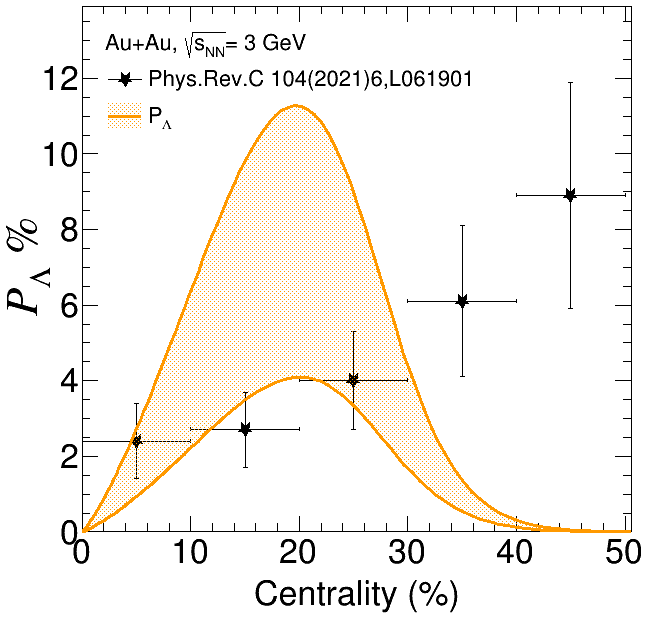}
 \caption{$\Lambda$ global polarization as a function of centrality calculated  for Au+Au collisions at $\sqrt{s_{NN}} = 3$ GeV with the number of $\Lambda$s estimated using Eq.~(\ref{eq:notheta}) compared with data from the STAR Co\-lla\-boration~\cite{STAR:2021beb}. The polarization increases for larger centrality.}
 \label{Fig:newpol}
\end{figure}


Figure~\ref{Fig:Nqgptheta}, shows
the number of $\Lambda$s calculated with a step function (continuous lines), compared with the ones calculated with Eq.~(\ref{eq:notheta}) for two different parameters (dashed lines). We observe that for $k=2$ it is possible to obtain $\Lambda$s from the core for larger impact parameter values. The production of $\Lambda$s for larger values of the impact parameter, increases the polarization for the $20-30$\% centrality bin and is non-vanishing for the $30-40$\% centrality bin. However, as shown in Fig.~\ref{Fig:newpol}, we are still a long way away from properly describing data in the most peripheral bins.

Another possible model improvement consists of including the polarization of $\Lambda$s created in the corona. Recall that in the model, $\Lambda$s produced in the corona are taken as not polarized. However, it is well-known that $\Lambda$s are in fact produced with a polarization in p+p collisions, which constitute the dominant kind of reactions in the corona. This polarization needs to be also accounted for in the model.

\section{Future work}

The core-corona model, hereby considered~\cite{Ayala:2020soy,Ayala:2021xrn}, assumes that the contribution to the global $\Lambda$ polarization from the corona was not significant. To account for a  $\Lambda$ polarization contribution from the corona, the first of Eqs.~(\ref{eq1-mod}) should be modified to read as
\begin{eqnarray}
\mathcal{P}^\Lambda=\frac{ \mathcal{P}^\Lambda_{REC} + z\frac{
N_{\Lambda\ {\mbox{\tiny{QGP}}}} }{N_{\Lambda\ {\mbox{\tiny{REC}}}}}}{ 1 + \frac{N_{\Lambda\ {\mbox{\tiny{QGP}}}}}{N_{\Lambda\ {\mbox{\tiny{REC}}}}}}.
\label{eq2-mod}
\end{eqnarray}

We observe that this contribution can be significant for energies smaller than the one corresponding to the crossing of the curves shown in Fig. ~\ref{Fig:AgAg2}. For higher energies, when the ratio $1 + \frac{N_{\Lambda\ {\mbox{\tiny{QGP}}}}}{N_{\Lambda\ {\mbox{\tiny{REC}}}}} \rightarrow 1$, the contribution tends to the polarization that is obtained in N + N collisions. From data on p+p collisions we know that the transverse polarization does not vanish~\cite{Blobel:1977ms,Jaeger:1974in,E690:2001otd} and in fact that the mean polarization in the energy range $10 \leq \sqrt{s_{NN}} \leq 63$ GeV is ${\mathcal{P}}=-0.38 \pm 0.06$ \cite{Panagiotou:1986zq}. We are thus implementing a way to include this polarization referred to the total angular momentum to estimate its contribution. This is work in progress and will be reported elsewhere.

\section{Summary}

We have shown that a two-component model describes the main characteristics of the $\Lambda$ and $\bar{\Lambda}$ polarization excitation function in semi-central heavy-ion collisions. The change in abundances of $\Lambda$s created in the core with respect to those created in the corona, as a function of collision energy, is responsible for the peaking of both polarizations for  $\sqrt{s_{NN}} \lesssim 10$ GeV.

The model considers a simple expansion scenario, whereby  the volume and lifetime of the QGP can be estimated to compute the relaxation time for the spin alignment of the quark (antiquark) that is then responsible for the $\Lambda$ ($\bar{\Lambda}$) spin alignment with the thermal vorticity. This in turn can be used to compute the intrinsic polarizations. The model provides a differentiated behavior for the global $\Lambda$ and $\bar{\Lambda}$ polarizations, that show maxima sensitive to the production mechanisms in the core and corona at HADES, NICA and RHIC energies. Current work to improve the model involves including a mechanism where $\Lambda$ polarization in peripheral collisions can be extracted from remnant transverse polarization of processes taking place in the corona.

\section{Acknowledgements}

Support for this work was received in part by UNAM DGAPA-PAPIIT grant number IG100322 and by Consejo Nacional de Ciencia y Tecnolog\'ia grant number A1-S-7655. METY is grateful for the hospitality of Perimeter Institute where part of this work was carried out and this research was also supported in part by the Simons Foundation through the Simons Foundation Emmy Noether Fellows Program at Perimeter Institute. Research at Perimeter Institute is supported in part by the Government of Canada and by the Province of Ontario.

\bibliography{ref-ADMT}

\end{document}